\documentclass[conference]{IEEEtran}
\IEEEoverridecommandlockouts
\usepackage{cite}
\usepackage{breakurl}
\usepackage[hyphens]{url}
\usepackage[breaklinks]{hyperref}

\usepackage{subfig}
\usepackage{amsmath,amssymb,amsfonts}
\usepackage{algorithmic}
\usepackage{graphicx}
\usepackage{textcomp}
\usepackage{multirow}
\usepackage[table]{xcolor}
\usepackage{threeparttable}
\usepackage{stfloats}
\usepackage[bottom]{footmisc}
\usepackage{listings}
\usepackage{color}
\usepackage[T1]{fontenc}
\usepackage{upquote}
\usepackage{tabularx}
\usepackage{tablefootnote}

\definecolor{dkgreen}{rgb}{0,0.6,0}
\definecolor{gray}{rgb}{0.5,0.5,0.5}
\definecolor{mauve}{rgb}{0.58,0,0.82}

\def\BibTeX{{\rm B\kern-.05em{\sc i\kern-.025em b}\kern-.08em
    T\kern-.1667em\lower.7ex\hbox{E}\kern-.125emX}}
\begin{document}

\title{Defect Prediction Using Stylistic Metrics\\
}

\author{\IEEEauthorblockN{Rafed Muhammad Yasir}
\IEEEauthorblockA{\textit{Institute of Information Technology} \\
\textit{University of Dhaka}\\
Dhaka, Bangladesh \\
bsse0733@iit.du.ac.bd}
\and
\and
\IEEEauthorblockN{Dr. Ahmedul Kabir}
\IEEEauthorblockA{\textit{Institute of Information Technology} \\
\textit{University of Dhaka}\\
Dhaka, Bangladesh \\
kabir@iit.du.ac.bd}
}

\maketitle

\textit{}
\begin{abstract}
Defect prediction is one of the most popular research topics due to its potentiality to minimize software quality assurance effort. Existing approaches have examined defect prediction from various perspective such as complexity and developer metrics. However, none of these consider programming style for defect prediction. This paper aims at analyzing the impact of stylistic metrics on both within-project and cross-project defect prediction. For prediction, 4 widely used machine learning algorithms namely Naive Bayes, Support Vector Machine, Decision Tree and Logistic Regression are used. The experiment is conducted on 14 releases of 5 popular, open source projects. F1, Precision and Recall are inspected to evaluate the results. Results reveal that stylistic metrics are good predictor of defects.
\end{abstract}


\section{Introduction}

A defect is a deviation between the expected and actual behavior of a program \cite{monperrus2018automatic}. Fixing defects can consume approximately 80\% of the total cost of a software project \cite{zhang2016cross}. To reduce this cost, defect prediction is required. It can help test managers to locate bugs and optimise the
allocation of limited testing resources \cite{song2018comprehensive}.

Existing studies in defect prediction can be divided into two categories namely within-project (both training and testing is performed on a single project) and cross-project (training and testing data are from different projects) defect prediction \cite{yan2017file}. 
Both of these categories have been examined from various viewpoints such as size and complexity metrics, code smells etc. However, the impact of programming style on defect prediction has not been analyzed yet. Programming style reveals individual's choice at the time of writing source code, such as preference of \textit{for} loop over \textit{while} loop \cite{mi2016measuring}. Oman et al. found that code complexity metrics are not enough to characterize programming style \cite{oman1989programming}. Therefore, stylistic metrics may improve defect prediction by capturing information which can not be covered by complexity metrics. 

This paper investigates the impact of stylistic metrics on both within-project and cross-project defect prediction. It uses 60 stylistic metrics, e.g., percentage of \textit{if-else}, and predicts whether a file is buggy or clean. For prediction, 4 widely used machine learning algorithms namely Naive Bayes (NB), Support Vector Machine (SVM), Decision Tree (DT), and Logistic Regression (LR) are used \cite{he2012investigation}, \cite{wang2016automatically}, \cite{soltanifar2016software}.

To evaluate the models, 14 releases of 5 popular, open source projects such as opencv, emscripten are used. 3 metrics namely F1, Recall and Precision are used for evaluation. Similar to \cite{he2012investigation}, models yielding Recall greater than 70\% and Precision greater than 50\% are considered as acceptable results. Among the 4 algorithms which obtains highest mean F1, is regarded as the best. For within-project prediction, Decision Tree provides the best result (mean F1 is 78.33). For cross-project prediction, Decision Tree and Support Vector Machine yield the best results (mean F1s are 72.07 and 72.57 respectively). Furthermore, 9 out of 14 releases in cross-project defect prediction obtain acceptable result. In within-project defect prediction, 6 out of 9 cases are acceptable result.

\section{Background and Related Work}
Defect prediction can be divided into two categories- within-project and cross-project defect prediction \cite{yan2017file}. If both training and testing is performed on a single project, it is called within-project defect prediction. On the other hand, when model is trained using data from one or several projects and evaluated on another project, it is called cross-project defect prediction. For new projects or projects with limited historical data, it is difficult to build within-project defect prediction model. For these projects, cross-project defect prediction is useful. However, the different distribution of defects in the training and test project makes cross-project defect prediction difficult.

For within-project and cross-project defect prediction, existing techniques have used mostly code and process metrics\cite{li2018progress}. Code metrics refer to source code properties such as size and complexity. Process metrics are collected from historical information archived in different software repositories (e.g., version control and issue tracking systems). For example, number of developers. \cite{jureczko2010using}, \cite{he2012investigation}, \cite{wang2016automatically} focused on code metrics, whereas \cite{matsumoto2010analysis}, \cite{soltanifar2016software} considered process metrics.

Jureczko and Spinellis used ckjm metrics suite (e.g., weighted methods per class, lines of code) for within-project defect prediction \cite{jureczko2010using}. They collected these metrics for five proprietary and eleven open source projects such as Lucene, Log4j etc. Each of the project had at least two versions. For each project, regression model is constructed using project version \textit{i-1} and tested on version \textit{i}. Their models were able to find 80\% of the defects within 36\% (mean value) of the classes.

He et al. used the same set of metrics as \cite{jureczko2010using} to analyze cross-project defect prediction \cite{he2012investigation}. They investigated 34 releases of 10 open source projects such as Ant, jEdit etc. They used 5 machine learning algorithms namely Naive Bayes, J48, Support Vector Machine, Decision Table, and Logistic Regression to build cross-project and within-project prediction models. Their study revealed that training data from carefully selected cross-project may provide better prediction results than one from the same project.

Wang et al. used semantic features, extracted from programs' Abstract Syntax Trees (ASTs) to improve defect prediction \cite{wang2016automatically}. For example, type of an AST node. They generated these features using Deep Belief Network. Usng these features, they built both within-project and cross-project defect prediction models. Results reveal that these semantic features significantly improve both within-project and cross-project defect prediction.

Matsumoto et al. investigated the effect of developers feature on defect prediction \cite{matsumoto2010analysis}. They considered two types of developer metrics that characterize developer\rq s activity (e.g., number of commitments) and modules revised by developers (e.g., number of developers revising module) respectively. Their experiment on 3 versions of Eclipse (version 3.00, 3.10 and 3.20) revealed that developer metrics can improve within-project defect prediction compared to models using static or change metrics.

Soltanifar et al. examined the impact of code smells (e.g., excessive parameter list) on within-project defect prediction \cite{soltanifar2016software}. For this purpose, two algorithms namely Naive Bayes and Logistic Regression were used. The developed models were compared with models using code churn metrics (e.g., lines of code added/deleted). The comparison was conducted on 2 projects from telecommunication industry and web content management system. Their study found that code smell is a good indicator of defect proneness.

The above discussion indicates that defect prediction has been analyzed from various viewpoints such as ckjm metrics, semantic metrics, code smells. However, none of these investigated the impact of stylistic metrics on defect prediction. Therefore, this paper initiates the work of analyzing within-project and cross-project defect prediction from stylistic metrics perspective.
\section{Methodology}

\subsection{Dataset}

To conduct experiments, 14 releases of 5 popular, open source (available on GitHub) c++ projects, such as, opencv, bitcoin, are used. The description of the projects are shown in Table I. For example, project opencv has 27709 commits and 40800 stars.

\noindent
\hspace*{\fill}
\bgroup
\def\arraystretch{1.5}%
\begin{table}[h]
\centering
\caption{Description of the Projects}
\begin{tabular}{|c|c|c|}
\hline
\textbf{Project} & \textbf{Number of Commits} & \textbf{Number of Stars} \\ \hline
bitcoin\tablefootnote{https://github.com/bitcoin/bitcoin}          & 22493                      & 41600                    \\ \hline
opencv\tablefootnote{https://github.com/opencv/opencv}           & 27709                      & 40800                    \\ \hline
rethinkdb\tablefootnote{https://github.com/rethinkdb/rethinkdb}        & 33422                      & 23000                    \\ \hline
emscripten\tablefootnote{https://github.com/emscripten-core/emscripten}       & 20271                      & 18300                    \\ \hline
cocos2d-x\tablefootnote{https://github.com/cocos2d/cocos2d-x}        & 37330                      & 14000                    \\ \hline
\end{tabular}
\end{table}
\egroup
\hspace*{\fill}
 
To prepare the dataset, the git repositories of each project are collected. The master branch is selected in all cases. If the master branch is not available, the default branch is selected. Next, each commit of the repositories are analyzed to identify whether it is a bug fixing commit. A commit message is searched for keywords such as "Fixed", "Bug", "Patch" and other expressions \cite{sliwerski2005changes} to determine if it is bug fixing. If a bug fixing commit is found, the bug introducing commit of the files fixed are found using the SZZ algorithm \cite{Spadini2018}. Next, the files between the bug introducing commit and bug fixing commit are labeled as buggy. All the other files are by default labeled as clean. While labeling, only the source files (cc, cxx, cpp, cu, c) are considered. For the labeled files of all versions, 60 stylistic features, proposed in \cite{mi2016measuring}, are extracted. For example, average number of functions and variable name length.

\noindent
\hspace*{\fill}
\bgroup
\def\arraystretch{1.5}%
\begin{table}[h]
\centering
\caption{Summary of the Dataset}
\begin{tabular}{|c|c|c|c|}
\hline
\textbf{Release} & \textbf{\begin{tabular}[c]{@{}c@{}}Total Number \\ of Files\end{tabular}} & \textbf{\begin{tabular}[c]{@{}c@{}}Number of \\ Buggy Files\end{tabular}} & \textbf{\begin{tabular}[c]{@{}c@{}}Percentage of \\ Buggy Files\end{tabular}} \\ \hline
bitcoin-0.7.0    & 90                                                                        & 64                                                                        & 71.11                                                                         \\ \hline
bitcoin-0.8.0    & 165                                                                       & 72                                                                        & 43.64                                                                         \\ \hline
cocos2d-x-3.0.0  & 751                                                                       & 362                                                                       & 48.20                                                                         \\ \hline
cocos2d-x-3.5.0  & 868                                                                       & 583                                                                       & 67.17                                                                         \\ \hline
cocos2d-x-3.8.0  & 993                                                                       & 664                                                                       & 66.87                                                                         \\ \hline
emscripten-1.25  & 1647                                                                      & 103                                                                       & 6.25                                                                          \\ \hline
emscripten-1.30  & 1890                                                                      & 136                                                                       & 7.20                                                                          \\ \hline
emscripten-1.35  & 2765                                                                      & 337                                                                       & 12.19                                                                         \\ \hline
opencv-2.4.6.2   & 1065                                                                      & 569                                                                       & 53.43                                                                         \\ \hline
opencv-2.4.7     & 1083                                                                      & 571                                                                       & 52.72                                                                         \\ \hline
opencv-3.0.0     & 1106                                                                      & 608                                                                       & 54.97                                                                         \\ \hline
rethinkdb-1.8    & 448                                                                       & 174                                                                       & 38.84                                                                         \\ \hline
rethinkdb-1.12   & 470                                                                       & 184                                                                       & 39.15                                                                         \\ \hline
rethinkdb-1.16   & 526                                                                       & 212                                                                       & 40.30                                                                         \\ \hline
\end{tabular}
\end{table}
\egroup
\hspace*{\fill}

Since change patterns in the first three years of a project are unstable, versions released in those years are excluded from the dataset \cite{jiang2013personalized}, \cite{tan2015online}. Furthermore, latest changes may be labeled clean even if they are buggy. Therefore, versions released in the recent three years of a project are excluded as well \cite{jiang2013personalized}, \cite{tan2015online}.

If there are more than three versions of a project after filtering, a maximum of three versions are selected, as followed in \cite{wang2016automatically}. The versions selected and the percentage of buggy files in each of them are listed in Table II.

\subsection{Approach}
This paper analyzes the impact of stylistic metrics on both within-project and cross-project defect prediction. For prediction, 4 machine learning algorithms namely Naive Bayes (NB), Support Vector Machine (SVM), Decision Tree (DT), and Logistic Regression (LR) are selected since they are widely used in defect prediction \cite{he2012investigation}, \cite{wang2016automatically}, \cite{soltanifar2016software}. Jiarpakdee et al. suggested to remove correlated metrics before constructing a defect model since they may impact the interpretation of the model \cite{8608002}. Therefore, stylistic metrics exceeding Variance Inflation Factor (VIF) score 5 are removed. This is a commonly used threshold in prior studies \cite{8608002}, \cite{fox2015applied}, \cite{mcintosh2014impact}. Table II reveals that the defect data is imbalanced. Hence, re-sampling technique SMOTE is used on the training data, as followed in \cite{wang2016automatically}.

\begin{figure}[htbp]
\centering
\centerline{\includegraphics[width=0.50\textwidth]{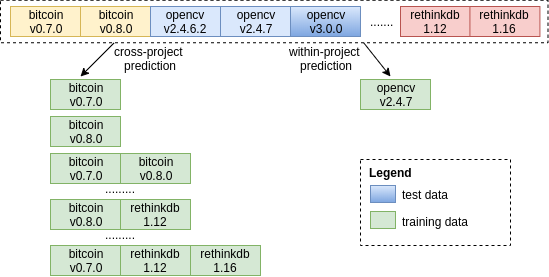}}
\caption{Training and Test Data Selection for Within-project and Cross-project Defect Prediction}
\label{worst}
\end{figure}

For within-project defect prediction, data from version \textit{i} and \textit{i+1} are used for training and testing the model respectively, as shown in Fig 1. For example, data of \textit{bitcoin-0.7.0} is used to predict defects in \textit{bitcoin-0.8.0}. For cross-project defect prediction, combination of all releases except the test project are used for training (shown in Fig 1). Similar to \cite{he2012investigation}, combinations upto 3 releases are used due to limited computing power. For example, to test \textit{bitcoin-0.7.0}, the training data are (\textit{cocos2d-x-3.0.0}), (\textit{opencv-2.4.7}), (\textit{cocos2d-x-3.0.0}, \textit{cocos2d-x-3.5.0}), (\textit{cocos2d-x-3.0.0}, \textit{emscripten-1.25}), (\textit{cocos2d-x-3.8.0}, \textit{opencv-3.0.0}, \textit{rethinkdb 1.16}) and so on.

\section{Experiment}
\subsection{Performance Evaluation}
In this study, Precision, Recall, and F1 are used to measure defect prediction results. These metrics are widely used for evaluation in defect prediction \cite{wang2016automatically}. These metrics are calculated using (1), (2) and (3).

\begin{equation} 
Precision = \frac{True\:Positive}{True\:Positive + False\:Positive}
\end{equation}

\begin{equation} 
Recall = \frac{True\:Positive}{True\:Positive + False\:Negative}
\end{equation}

\begin{equation} 
F1 = \frac{2 * Precision * Recall}{Precision + Recall}
\end{equation}

True positive is the number of buggy files that are classified correctly. False positive and false negative are the numbers of files wrongly classified as buggy and clean respectively. The values of Recall and Precision are usually mutually exclusive \cite{he2012investigation}. This paper considers prediction results with Recall greater than 70\% and Precision greater than 50\% as acceptable results as followed in \cite{he2012investigation}. F1 outputs a single value by considering both precision and recall. The higher the F value, the better the prediction result is. 

\subsection{Result Analysis}
Table III shows the result obtained using the four algorithms. To assess prediction results, mean values of F1, Precision and Recall are calculated.

\noindent
\hspace*{\fill}
\bgroup
\def\arraystretch{1.5}%
\begin{table}[h]
\caption{Prediction Result of the Four Algorithms}
\begin{tabular}{|c|c|c|c|c|c|}
\hline
\textbf{Category}               & \textbf{Indicators} & \textbf{NB} & \textbf{DT} & \textbf{SVM} & \textbf{LR} \\ \hline
\multirow{3}{*}{Within-project} & Mean(F1)            & 59.33       & 78.33       & 63.44        & 63.78       \\ \cline{2-6} 
                                & Mean(Precision)     & 74.22       & 83.33       & 70.33        & 72.44       \\ \cline{2-6} 
                                & Mean(Recall)        & 52.67       & 77.11       & 61.78        & 57.33       \\ \hline
\multirow{3}{*}{Cross-project}  & Mean(F1)            & 69.93       & 72.07       & 72.57        & 71.64       \\ \cline{2-6} 
                                & Mean(Precision)     & 63.14       & 63.64       & 65.14        & 64.43       \\ \cline{2-6} 
                                & Mean(Recall)        & 82.57       & 86.64       & 86.43        & 88.14       \\ \hline
\end{tabular}
\end{table}
\egroup
\hspace*{\fill}

\noindent
\hspace*{\fill}
\bgroup
\def\arraystretch{1.5}%
\begin{table}[h]
\centering
\caption{Within-Project Defect Prediction Result Using Decision Tree}
\begin{tabular}{|c|c|c|c|c|}
\hline
\textbf{Project} & \textbf{Releases (Tr -\textgreater{} T)} & \textbf{Precision} & \textbf{Recall} & \textbf{F1}    \\ \hline
Bitcoin          & 0.7.0 -\textgreater{} 0.8.0              & 90                 & 28              & 43             \\ \hline
Cocos2d-x        & 3.0.0 -\textgreater{} 3.5.0              & 50                 & 58              & 53             \\ \hline
*Cocos2d-x        & 3.5.0 -\textgreater{} 3.8.0              & 79                 & 81              & 80             \\ \hline
*Emscripten       & 1.25 -\textgreater{} 1.30                & 98                 & 98              & 98             \\ \hline
*Emscripten       & 1.30 -\textgreater{} 1.35                & 94                 & 96              & 95             \\ \hline
*Opencv           & 2.4.6.2 -\textgreater{} 2.4.7            & 95                 & 96              & 96             \\ \hline
Opencv           & 2.4.7 -\textgreater{} 3.0.0              & 72                 & 67              & 69             \\ \hline
*RethinkDB        & 1.8 -\textgreater{} 1.12                 & 87                 & 90              & 89             \\ \hline
*RethinkDB        & 1.12 -\textgreater{} 1.14                & 85                 & 80              & 82             \\ \hline
\multicolumn{2}{|c|}{\textbf{Mean}}                       & \textbf{83.33}     & \textbf{77.11}  & \textbf{78.33} \\ \hline
\end{tabular}
\end{table}
\egroup
\hspace*{\fill}

For within-project defect prediction, Decision Tree performed the best, based on mean F1 (78.33). Wilcoxon Signed-Rank test is performed to check whether this result is statistically significant \cite{walpole1993probability}. The mean F1 is significantly greater for Decision Tree, at significance level = 0.05. The mean values of Precision and Recall for Decision Tree are 83.33 and 77.11 respectively. Detailed results for Decision Tree are presented in Table IV. In the table, 6 out of 9 cases are acceptable result (Recall greater than 70\% and Precision greater than 50\%), which are shown using star(*) mark.

\noindent
\hspace*{\fill}
\bgroup
\def\arraystretch{1.8}%
\begin{table*}[t]
\caption{Cross-Project Defect Prediction Result Using Support Vector Machine}
\centering
\begin{tabular}{|c|c|c|c|c|c|c|}
\hline
\multirow{2}{*}{\textbf{Release}} & \multicolumn{3}{c|}{\textbf{Most suitable training set}}  & \multirow{2}{*}{\textbf{Precision}} & \multirow{2}{*}{\textbf{Recall}} & \multirow{2}{*}{\textbf{F}} \\ \cline{2-4}
                                  & \textbf{Release1} & \textbf{Release2} & \textbf{Release3} &                                     &                                  &                             \\ \hline
bitcoin-0.7.0                       & emscripten-1.35   & opencv-3.0        & rethinkdb-1.8     & 50                                  & 62                               & 55                          \\ \hline
*bitcoin-0.8.0                       & cocos2d-x-3.0     & emscripten-1.35   & opencv-2.4.7      & 79                                  & 80                               & 79                          \\ \hline
*cocos2d-x-3.0                     & opencv-3.0        & rethinkdb-1.8     & rethinkdb-1.16    & 80                                  & 72                               & 76                          \\ \hline
cocos2d-x-3.5.0                     & emscripten-1.30   & rethinkdb-1.12    & rethinkdb-1.16    & 33                                  & 99                               & 50                          \\ \hline
cocos2d-x-3.8.0                     & bitcoin-0.7.0       & bitcoin-0.8.0       & rethinkdb-1.16    & 53                                  & 47                               & 50                          \\ \hline
*emscripten-1.25                   & bitcoin-0.8.0       & cocos2d-x-3.0     & opencv-2.4.7      & 94                                  & 99                               & 96                          \\ \hline
*emscripten-1.30                   & bitcoin-0.8.0       & cocos2d-x-3.0     & opencv-2.4.7      & 93                                  & 100                              & 96                          \\ \hline
*emscripten-1.35                   & cocos2d-x-3.0     & opencv-2.4.6.2    & -                 & 88                                  & 99                               & 93                          \\ \hline
*opencv-2.4.6.2                    & bitcoin-0.8.0       & emscripten-1.35   & rethinkdb-1.12    & 52                                  & 87                               & 65                          \\ \hline
opencv-2.4.7                      & emscripten-1.25   & -                 & -                 & 50                                  & 93                               & 65                          \\ \hline
opencv-3.0                        & cocos2d-x-3.5.0     & emscripten-1.25   & emscripten-1.35   & 49                                  & 88                               & 63                          \\ \hline
*rethinkdb-1.8                     & bitcoin-0.8.0       & emscripten-1.35   & opencv-3.0        & 62                                  & 97                               & 76                          \\ \hline
*rethinkdb-1.12                    & bitcoin-0.8.0       & cocos2d-x-3.0.0     & cocos2d-x-3.5.0     & 63                                  & 98                               & 76                          \\ \hline
*rethinkdb-1.16                    & bitcoin-0.8.0       & cocos2d-x-3.5.0     & emscripten-1.25   & 66                                  & 89                               & 76                          \\ \hline
\multicolumn{4}{|c|}{\textbf{Mean}}                                                           & \textbf{65.14}                      & \textbf{86.43}                   & \textbf{72.57}              \\ \hline
\end{tabular}
\end{table*}
\egroup
\hspace*{\fill}

For cross-project defect prediction, Support Vector Machine performed the best, based on mean F1 (72.57). Wilcoxon Signed-Rank test reveal that the mean F1 of Support Vector Machine is significantly greater than Naive Bayes and Logistic Regression. However, there is no significant difference between Support Vector Machine and Decision Tree. Detailed results for Support Vector Machine are presented in Table IV. Here, 9 out of 14 releases obtain Recall greater than 70\% and Precision greater than 50\%. The result may be further improved by combining more than 3 project data for training.

\section{Threats to Validity}
This section presents potential aspects which may threat the validity of the study:

\begin{itemize}

\item \textbf{External Validity:} The experiment is conducted on 14 releases of 5 popular, open source, c++ projects. The findings may not be generally applicable for other languages such as Java or other projects. 

\item \textbf{Construct Validity:} Following prior study \cite{he2012investigation}, the thresholds of Recall (greater than 70\%) and Precision (greater than 50\%) for acceptable result are set. Next, 4 widely used machine learning algorithms are chosen in this study. Using other algorithms may impact the results. For capturing programming style, 60 metrics proposed by \cite{mi2016measuring} are used. These metrics may not be enough to completely characterize programming styles. 

\item \textbf{ Reliability:} To ensure reliability of the study, the dataset is made publicly available at
\url{https://doi.org/10.5281/zenodo.3591973}.
\end{itemize}

\section{Conclusion}
This paper investigates the impact of stylistic features on defect prediction. The study is conducted for both within-project and cross-project defect prediction. 4 machine learning models namely Naive Bayes, Logistic Regression, Support Vector Machine and Decision Tree are trained. F1, Precision and Recall are used to evaluate the results. Results show that stylistic features are good indicators of defects at a file level granularity. For within-project prediction, Decision Tree provides the best result (mean F1 is 78.33). For cross-project prediction, Decision Tree and Support Vector Machine yield the best results (mean F1s are 72.07 and 72.57 respectively).

In future, more than three combinations of training dataset for cross-project prediction will be analyzed to improve results. Furthermore, important stylistic metrics will be identified and incorporated with traditional metrics used in defect prediction such as OOP metrics and code churn metrics.

\section{Acknowledgement}
 This research is supported by Bangladesh Research and Education Network (BdREN).
\bibliography{mybib}

\begin{thebibliography}{10}

\bibitem{monperrus2018automatic}
M.~Monperrus, ``Automatic software repair: a bibliography,'' {\em ACM Computing
  Surveys (CSUR)}, vol.~51, no.~1, p.~17, 2018.

\bibitem{zhang2016cross}
F.~Zhang, Q.~Zheng, Y.~Zou, and A.~E. Hassan, ``Cross-project defect prediction
  using a connectivity-based unsupervised classifier,'' in {\em Proceedings of
  the 38th International Conference on Software Engineering}, pp.~309--320,
  ACM, 2016.

\bibitem{song2018comprehensive}
Q.~Song, Y.~Guo, and M.~Shepperd, ``A comprehensive investigation of the role
  of imbalanced learning for software defect prediction,'' {\em IEEE
  Transactions on Software Engineering}, 2018.

\bibitem{yan2017file}
M.~Yan, Y.~Fang, D.~Lo, X.~Xia, and X.~Zhang, ``File-level defect prediction:
  Unsupervised vs. supervised models,'' in {\em 2017 ACM/IEEE International
  Symposium on Empirical Software Engineering and Measurement (ESEM)},
  pp.~344--353, IEEE, 2017.

\bibitem{mi2016measuring}
Q.~Mi, J.~Keung, and Y.~Yu, ``Measuring the stylistic inconsistency in software
  projects using hierarchical agglomerative clustering,'' in {\em Proceedings
  of the The 12th International Conference on Predictive Models and Data
  Analytics in Software Engineering}, p.~5, ACM, 2016.

\bibitem{oman1989programming}
P.~W. Oman and C.~R. Cook, ``Programming style authorship analysis,'' in {\em
  Proceedings of the 17th conference on ACM Annual Computer Science
  Conference}, pp.~320--326, ACM, 1989.

\bibitem{he2012investigation}
Z.~He, F.~Shu, Y.~Yang, M.~Li, and Q.~Wang, ``An investigation on the
  feasibility of cross-project defect prediction,'' {\em Automated Software
  Engineering}, vol.~19, no.~2, pp.~167--199, 2012.

\bibitem{wang2016automatically}
S.~Wang, T.~Liu, and L.~Tan, ``Automatically learning semantic features for
  defect prediction,'' in {\em 2016 IEEE/ACM 38th International Conference on
  Software Engineering (ICSE)}, pp.~297--308, IEEE, 2016.

\bibitem{soltanifar2016software}
B.~Soltanifar, S.~Akbarinasaji, B.~Caglayan, A.~B. Bener, A.~Filiz, and B.~M.
  Kramer, ``Software analytics in practice: a defect prediction model using
  code smells,'' in {\em Proceedings of the 20th International Database
  Engineering \& Applications Symposium}, pp.~148--155, ACM, 2016.

\bibitem{li2018progress}
Z.~Li, X.-Y. Jing, and X.~Zhu, ``Progress on approaches to software defect
  prediction,'' {\em IET Software}, vol.~12, no.~3, pp.~161--175, 2018.

\bibitem{jureczko2010using}
M.~Jureczko and D.~Spinellis, ``Using object-oriented design metrics to predict
  software defects,'' {\em Models and Methods of System Dependability. Oficyna
  Wydawnicza Politechniki Wroc{\l}awskiej}, pp.~69--81, 2010.

\bibitem{matsumoto2010analysis}
S.~Matsumoto, Y.~Kamei, A.~Monden, K.-i. Matsumoto, and M.~Nakamura, ``An
  analysis of developer metrics for fault prediction,'' in {\em Proceedings of
  the 6th International Conference on Predictive Models in Software
  Engineering}, p.~18, ACM, 2010.

\bibitem{sliwerski2005changes}
J.~{\'S}liwerski, T.~Zimmermann, and A.~Zeller, ``When do changes induce
  fixes?,'' in {\em ACM sigsoft software engineering notes}, vol.~30, pp.~1--5,
  ACM, 2005.

\bibitem{Spadini2018}
D.~Spadini, M.~Aniche, and A.~Bacchelli, ``{PyDriller: Python framework for
  mining software repositories},'' in {\em Proceedings of the 2018 26th ACM
  Joint Meeting on European Software Engineering Conference and Symposium on
  the Foundations of Software Engineering - ESEC/FSE 2018}, (New York, New
  York, USA), pp.~908--911, ACM Press, 2018.

\bibitem{jiang2013personalized}
T.~Jiang, L.~Tan, and S.~Kim, ``Personalized defect prediction,'' in {\em 2013
  28th IEEE/ACM International Conference on Automated Software Engineering
  (ASE)}, pp.~279--289, Ieee, 2013.

\bibitem{tan2015online}
M.~Tan, L.~Tan, S.~Dara, and C.~Mayeux, ``Online defect prediction for
  imbalanced data,'' in {\em 2015 IEEE/ACM 37th IEEE International Conference
  on Software Engineering}, vol.~2, pp.~99--108, IEEE, 2015.

\bibitem{8608002}
J.~{Jiarpakdee}, C.~{Tantithamthavorn}, and A.~E. {Hassan}, ``The impact of
  correlated metrics on the interpretation of defect models,'' {\em IEEE
  Transactions on Software Engineering}, pp.~1--1, 2019.

\bibitem{fox2015applied}
J.~Fox, {\em Applied regression analysis and generalized linear models}.
\newblock Sage Publications, 2015.

\bibitem{mcintosh2014impact}
S.~McIntosh, Y.~Kamei, B.~Adams, and A.~E. Hassan, ``The impact of code review
  coverage and code review participation on software quality: A case study of
  the qt, vtk, and itk projects,'' in {\em Proceedings of the 11th Working
  Conference on Mining Software Repositories}, pp.~192--201, ACM, 2014.

\bibitem{walpole1993probability}
R.~E. Walpole, R.~H. Myers, S.~L. Myers, and K.~Ye, {\em Probability and
  statistics for engineers and scientists}, vol.~5.
\newblock Macmillan New York, 1993.

\end{thebibliography}
\bibliographystyle{ieeetr}

\end{document}